\documentclass[twocolumn,preprintnumbers,amsmath,aps]{revtex4}
\usepackage{graphicx}
\usepackage{dcolumn}
\usepackage{bm}
\usepackage{color}
\begin{document}
\small
\newcommand{\kvec}{\mbox{{\scriptsize {\bf k}}}}
\newcommand{\qvec}{\mbox{{\scriptsize {\bf q}}}}
\def\eq#1{(\ref{#1})}
\def\fig#1{\ref{#1}}
\def\tab#1{\ref{#1}}
\title{
Characteristics of the Eliashberg formalism on the example of high-pressure superconducting state in phosphor}
\author{A. M. Duda$^{\left(1\right)}$}
\email{aduda@wip.pcz.pl}
\author{R. Szcz{\c{e}}{\'s}niak$^{\left(1,2\right)}$}
\email{szczesni@wip.pcz.pl} 
\author{M. A. Sowi{\'n}ska$^{\left(1\right)}$} 
\author{I. A. Domagalska$^{\left(1\right)}$}
\affiliation{$^1$ Institute of Physics, Cz{\c{e}}stochowa University of Technology, Ave. Armii Krajowej 19, 42-200 Cz{\c{e}}stochowa, Poland}
\affiliation{$^2$ Institute of Physics, Jan D{\l}ugosz University in Cz{\c{e}}stochowa, Ave. Armii Krajowej 13/15, 42-200 Cz{\c{e}}stochowa, Poland}
\date{\today} 
\begin{abstract}
The work describes the properties of the high-pressure superconducting state in phosphor: $p\in\{20, 30, 40, 70\}$ GPa. 
The calculations were performed in the framework of the Eliashberg formalism, which is the natural generalization of the BCS theory. 
The exceptional attention was paid to the accurate presentation of the used analysis scheme. 
With respect to the superconducting state in phosphor it was shown that: 
(i) the observed not-high values of the critical temperature ($\left[T_{C}\right]_{p=30{\rm GPa}}^{\rm max}=8.45$ K) result not only from the low values of the electron - phonon coupling constant, but also from the very strong depairing Coulomb interactions, (ii) the inconsiderable strong - coupling and retardation effects force the dimensionless ratios $R_{\Delta}$, $R_{C}$, and $R_{H}$ - related to the critical temperature, 
the order parameter, the specific heat and the thermodynamic critical field - to take the values close to the BCS predictions.\\

\end{abstract}
\maketitle

\section{Hamiltonian and fundamental equations of BCS model and Eliashberg formalism}

The first microscopic theory of the superconducting state was formulated in 1957 by Bardeen, Cooper and Schrieffer (the so-called "BCS model") \cite{Bardeen1957A}, \cite{Bardeen1957B}. In the framework of the method of the second quantization the BCS Hamiltonian can be written with the following formula \cite{Fetter1971A}, \cite{Elk1979A}: 
\begin{equation}
\label{r1}
H=\sum_{\kvec\sigma} \varepsilon_{\kvec} c_{\kvec\sigma}^{\dag} c_{\kvec\sigma} - V\sum_{\kvec\kvec'}{^{'}} c_{\kvec\uparrow}^{\dag} c_{-\kvec\downarrow}^{\dag} c_{-\kvec'\downarrow} c_{\kvec'\uparrow}, 
\end{equation}
where the function $\varepsilon_{\kvec}$ represents the electron band energy, $V$ is the effective pairing potential, which value is determined by the matrix elements of the electron - phonon interaction, the electron band energy and the phonon energy. 
The symbols $c_{\kvec\sigma}^{\dag}$ and $c_{\kvec\sigma}$ represent the creation and annihilation operator of the electron state in the momentum representation ({\bf k}) for the spin $\sigma\in\{\uparrow,\downarrow\}$. It should be noted that the sum denoted by the sign $^{'}$ ought to be calculated only for those values of the momentums, for which the condition: 
$-\Omega_{\rm max}<\varepsilon_{\kvec}<\Omega_{\rm max}$ is fulfilled, where $\Omega_{\rm max}$ represents the Debye energy. 
In the considered case the effective pairing potential is positive, which allows the formation of the superconducting condensate. The fundamental equation of the BCS theory for the order parameter 
($\Delta \equiv V\sum_{\kvec}{^{'}} \langle c_{-\kvec\downarrow} c_{\kvec\uparrow}\rangle$) is derived directly from the Hamiltonian (\ref{r1}) using the mean field approximation to the interaction term. As a result the following can be obtained: 
\begin{equation}
\label{r2}
1 = V\sum_{\kvec}{^{'}}\frac{1}{2\sqrt{\varepsilon_{\kvec}^{2}+|\Delta(T)|^{2}}} \tanh\frac{\sqrt{\varepsilon_{\kvec}^{2}+|\Delta(T)|^{2}}}{2k_BT},
\end{equation}
where $k_{B}$ is the Boltzmann constant. Let us notice that the equation (\ref{r2}) cannot be solved analytically. However, in the limit cases: 
$T\rightarrow T_{C}$ and $T\rightarrow 0$ K the relatively simple calculations allow us to obtain the formulas for the critical temperature and the value of the order parameter:
$k_BT_C = 1.13\Omega_{\rm max}\exp\left[-1/\lambda\right]$,
$\Delta\left(0\right)=2\Omega_{\rm max}\exp\left[-1/\lambda\right]$,
where the electron-phonon coupling constant ($\lambda$) in the BCS model is given by: 
$\lambda\equiv\rho\left(0\right)V$ (the quantity $\rho\left(0\right)$ represents the electron density of states on the Fermi surface).
The BCS theory predicts the existence of the universal thermodynamic ratios, which are defined below:
\begin{equation}
\label{r3}
R_{\Delta}\equiv\frac{2\Delta\left(0\right)}{k_BT_C}=3.53,
\end{equation}
\begin{equation}
\label{r4}
R_C\equiv\frac{C^{S}\left(T_C\right)-C^{N}\left(T_C\right)}{C^N\left(T_C\right)}=1.43,
\end{equation}
and
\begin{equation}
\label{r5}
R_H\equiv\frac{T_CC^N\left(T_C\right)}{H_C^2\left(0\right)}=0.168.
\end{equation}
The symbols appearing in the formulas (\ref{r4}) and (\ref{r5}) denote respectively: ($C^S$) the specific heat of the superconducting state, 
($C^N$) the specific heat of the normal state, and ($H_C$) the thermodynamic critical field. It should be noted that the predictions of the BCS theory quantitatively agree with the experimental data only within the scope of the weak electron - phonon coupling ($\lambda\leq 0.3$).  

The Eliashberg formalism is the natural generalization of the BCS model (explicitly complies with the electron - phonon interaction). 
The starting point of the theory is the Hamiltonian, which models the linear coupling between the electron and the phonon sub-system \cite{Frohlich1950A}, \cite{Frohlich1954A}:
\begin{eqnarray}
\label{r6}
H&=&\sum_{\kvec\sigma}\varepsilon_{\kvec}c^{\dag}_{\kvec\sigma}c_{\kvec\sigma}
+\sum_{\qvec}\omega_{\qvec}b^{\dag}_{\qvec}b_{\qvec}\\ \nonumber
&+&\sum_{\kvec\sigma}g_{\kvec,\kvec+\qvec}c^{\dag}_{\kvec+\qvec\sigma}c_{\kvec\sigma}\left(b^{\dag}_{-\qvec}+b_{\qvec}\right).
\end{eqnarray}
The symbol $\omega_{\qvec}$ represents the phonon energy, $g_{\kvec,\kvec+\qvec}$ are the matrix elements of the electron - phonon interaction. The operator $b^{\dag}_{\qvec}$ ($b_{\qvec}$) creates (annihilates) the phonon state with the momentum ${\bf q}$. Based on the operator (\ref{r6}) and using the formalism of the thermodynamic Green functions of the Matsubara type it is possible to derive the Eliashberg equations on the imaginary axis ($i\equiv\sqrt{-1}$) \cite{Eliashberg1960A}, \cite{Carbotte1990A}:
\begin{eqnarray}
\label{r7}
\Delta_{n}Z_{n}=\pi k_{B}T\sum^{M}_{m=-M}
\frac{[K\left(i\omega_{n}-i\omega_{m}\right)-\mu^{\star}\left(\omega_{m}\right)]}{\sqrt{\omega^{2}_{m}+\Delta_{m}^2}}{\Delta_{m}},
\end{eqnarray}
and
\begin{eqnarray}
\label{r8}
Z_{n}=1+\pi k_{B}T\sum^{M}_{m=-M}
\frac{K\left(i\omega_{n}-i\omega_{m}\right)}{\sqrt{\omega^{2}_{m}+\Delta_{m}^{2}}}\frac{\omega_{m}}{\omega_{n}}Z_{m}.
\end{eqnarray}
The quantity $\Delta_{n}\equiv\Delta\left(i\omega_{n}\right)$ denotes the order parameter, while $Z_{n}\equiv Z\left(i\omega_{n}\right)$ is the wave function renormalization factor. The symbol $\omega_{n}$ is the fermion Matsubara frequency:  
$\omega_{n}\equiv\pi k_{B}T\left(2n-1\right)$. The values of the pairing kernel should be calculated on the basis of the formula:
$K\left(z\right)\equiv 2\int_0^{\Omega_{\rm max}}d\Omega\frac{\Omega}{\Omega ^2-z^{2}}\alpha^{2}F\left(\Omega\right)$.
The symbol $\alpha^{2}F\left(\Omega\right)$ represents the so-called Eliashberg function, which quantitatively models the electron - phonon interaction. The function $\alpha^{2}F\left(\Omega\right)$ can be determined either by using the data from the tunnel experiment or by referring to the results of the calculations from the first principles. In the classical Eliashberg formalism the depairing electron correlations were included in the parametric manner: 
$\mu^{\star}\left(\omega_m\right)\equiv\mu^{\star}\theta \left(\omega_{C}-|\omega_m|\right)$, whereas $\mu^{\star}$ is called the Coulomb pseudopotential, the symbol $\theta$ is the Heaviside function, and $\omega_{C}$ denotes the cut-off frequency, which value is usually several times higher than the value of $\Omega_{\rm{max}}$. The Eliashberg equations can be also derived in the case, in which the fitting parameter 
$\mu^{\star}$ is not present (mathematically this is a very tough issue). For this purpose, both the extended Hubbard Hamiltonian \cite{Spalek2015A} and the method of the analysis discussed in the paper \cite{Spalek2014A} should be used. The Eliashberg equations on the imaginary axis allow us to precisely calculate the critical temperature and the free energy difference between the superconducting and the normal state. They cannot, however, be used to determine the exact physical values of the order parameter and the effective mass of the electron. For this purpose, the Eliashberg equations should be analytically extended in such a way that the equations in the mixed representation can be obtained (determined simultaneously on the imaginary and real axis) \cite{Marsiglio1988A}:
\begin{widetext}
\begin{eqnarray}
\label{r9}
\phi\left(\omega+i\delta\right)&=&
                                  {\pi}{k_{B}T}\sum_{m=-M}^{M}
                                  \left[K\left(\omega-i\omega_{m}\right)-\mu^{\star}\left(\omega_{m}\right)\right]
                                  \frac{\phi_{m}}
                                  {\sqrt{\omega_m^2 Z^{2}_{m}+\phi^{2}_{m}}}\\ \nonumber
                              &+&i\pi\int_{0}^{+\infty}d\omega^{'}\alpha^{2}F\left(\omega^{'}\right)
                                  \Bigg[\left[f_{\rm BE}\left(\omega^{'}\right)+f_{\rm FD}\left(\omega^{'}-\omega\right)\right]
                                  \frac{\phi\left(\omega-\omega^{'}+i\delta\right)}
                                  {\sqrt{\left(\omega-\omega^{'}\right)^{2}Z^{2}\left(\omega-\omega^{'}+i\delta\right)
                                  -\phi^{2}\left(\omega-\omega^{'}+i\delta\right)}}\Bigg]\\ \nonumber
                              &+&i\pi\int_{0}^{+\infty}d\omega^{'}\alpha^{2}F\left(\omega^{'}\right)
                                  \Bigg[\left[f_{\rm BE}\left(\omega^{'}\right)+f_{\rm FD}\left(\omega^{'}+\omega\right)\right]
                                  \frac{\phi\left(\omega+\omega^{'}+i\delta\right)}
                                  {\sqrt{\left(\omega+\omega^{'}\right)^{2}Z^{2}\left(\omega+\omega^{'}+i\delta\right)
                                  -\phi^{2}\left(\omega+\omega^{'}+i\delta\right)}}\Bigg],
\end{eqnarray}
\begin{eqnarray}
\label{r10}
Z\left(\omega+i\delta\right)&=&
                                  1+\frac{i}{\omega}{\pi}{k_{B}T}\sum_{m=-M}^{M}
                                  K\left(\omega-i\omega_{m}\right)
                                  \frac{\omega_{m}Z_{m}}
                                  {\sqrt{\omega_m^2Z^{2}_{m}+\phi^{2}_{m}}}\\ \nonumber
                              &+&\frac{i\pi}{\omega}\int_{0}^{+\infty}d\omega^{'}\alpha^{2}F\left(\omega^{'}\right)
                                  \Bigg[\left[f_{\rm BE}\left(\omega^{'}\right)+f_{\rm FD}\left(\omega^{'}-\omega\right)\right]
                                 \frac{\left(\omega-\omega^{'}\right)Z\left(\omega-\omega^{'}+i\delta\right)}
                                  {\sqrt{\left(\omega-\omega^{'}\right)^{2}Z^{2}\left(\omega-\omega^{'}+i\delta\right)
                                  -\phi^{2}\left(\omega-\omega^{'}+i\delta\right)}}\Bigg]\\ \nonumber
                              &+&\frac{i\pi}{\omega}\int_{0}^{+\infty}d\omega^{'}\alpha^{2}F\left(\omega^{'}\right)
                                  \Bigg[\left[f_{\rm BE}\left(\omega^{'}\right)+f_{\rm FD}\left(\omega^{'}+\omega\right)\right]
                                  \frac{\left(\omega+\omega^{'}\right)Z\left(\omega+\omega^{'}+i\delta\right)}
                                  {\sqrt{\left(\omega+\omega^{'}\right)^{2}Z^{2}\left(\omega+\omega^{'}+i\delta\right)
                                  -\phi^{2}\left(\omega+\omega^{'}+i\delta\right)}}\Bigg],
\end{eqnarray}
\end{widetext}
where the symbols $f_{BE}\left(\omega\right)$ and $f_{FD}\left(\omega\right)$ represent respectively the Bose - Einstein function and 
the Fermi - Dirac function. Note that the order parameter is defined as follows:
$\Delta\left(\omega\right)\equiv\phi\left(\omega\right)\slash Z\left(\omega\right)$.

In the remaining part of the work, we have described the way of using the Eliashberg formalism to determine the characteristics of the superconducting state induced by the electron - phonon interaction. All relevant issues have been discussed on the example of the superconducting condensate in phosphor, which was subjected to the influence of the high pressure.  

\section{Superconducting phase in phosphor: the state of knowledge}
\begin{figure} 
\includegraphics[width=\columnwidth]{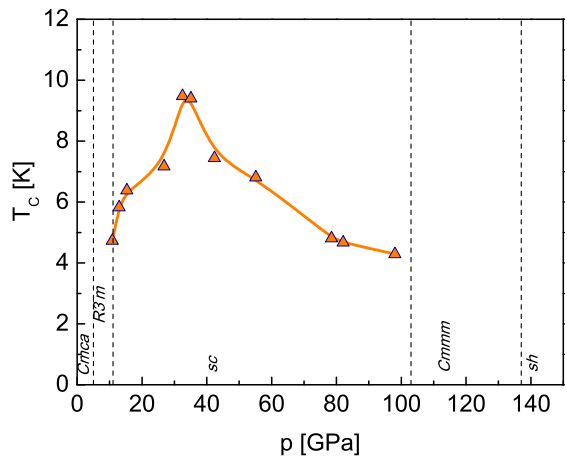}
\caption{\label{f1}
      The influence of the pressure on the value of the critical temperature in phosphor \cite{Karuzawa2002A}.}
\end{figure}
Five structural phase transitions can be observed in phosphor in the range of the pressure from normal to 262 GPa. In the normal conditions the black phosphor has the structure {\it Cmca}, which is stable up to the value of the pressure at 5 GPa \cite{Keyse1953A}, \cite{Morita1986A}. The existence of the structure ${\rm R\overline{3}m}$ proven above, vanishes at the pressure of 11.1 GPa, going into the metallic phase (sc) \cite{Jamieson1963A}. In the range from 103 GPa to 137 GPa the structure {\it Cmmm} is stable \cite{Akahama1999A}. Further, up to the pressure at 262 GPa, a simple structure sh has been observed. Whereas, for the pressure at $p>262$ GPa, the stability of the structure bcc has been noticed \cite{Akahama1999A}, \cite{Akahama2000A}.   
The superconducting state in phosphor was observed for the first time about fifty years ago \cite{Wittig1968A}, \cite{Berman1968A}. It should be noted, however, that as of today the exact dependence of the critical temperature on the pressure is not fully known because the changes of the values of $T_{C}$ are strongly correlated with the road passed on the $p$-$T$ diagram \cite{Kawamura1984A}, \cite{Kawamura1985A}. The results from 1985 suggest that the value of the critical temperature is equal to about 6 K (the structure sc) and weakly depends on the pressure \cite{Kawamura1985A}. On the other hand, the experimental data included in the work \cite{Wittig1985A} indicate the existence of two characteristic peaks in the course of the function $T_{C}\left(p\right)$, while the highest value of the critical temperature is about 10 K ($p=23$ GPa). The latest results are quite different and suggest the existence of the single maximum of the critical temperature ($T_{C}=9.5$ K) located in the vicinity of the pressure at $32$ GPa, which is presented in Figure \fig{f1} \cite{Karuzawa2002A}. Referring to the theoretical predictions, it should be noted that in general they are obtained by the use of the significant approximations. For example, in the paper \cite{Rajagopalan1989A} there is the evidence of the maximum $T_{C}$ located near the second experimental maximum of the critical temperature indicated in the publication \cite{Wittig1985A}. However, this theoretical work completely left out the effect of the pressure on the phonon spectrum. A similar approach in the work \cite{Aoki1987A} caused on incorrect definition of the function 
$T_{C}\left(p\right)$. Relatively new results \cite{Nagara2010A} suggest a very weak dependence of the critical temperature on the pressure - which is in agreement with the experimental data presented in \cite{Kawamura1985A}. However, this is in a sharp contrast with the recent experimental results \cite{Karuzawa2002A}. At this point, let us mention the doctoral thesis of Nixon \cite{Nixon2010A}, which expects the growth of the critical temperature ($T_{C}\in\left<8.5,11\right>$ K) together with the growing pressure ($p\in\left<10,35\right>$ GPa). However, the determined values of the Debye temperature are based only on the bulk modulus. 

\section{Thermodynamics of high-pressure superconducting state in phosphor: the Eliashberg formalism}

When analyzing the properties of the superconducting state in phosphor, we have taken into account the values of the pressure equal to: $20$ GPa, $30$ GPa, $40$ GPa, and $70$ GPa respectively, while the Eliashberg functions were determined in \cite{Chan2013A}. It is worth noting that the required calculations for the electron band structure, the phonon spectrum, and the electron - phonon interaction have been conducted in the full 
{\it ab initio} scheme. Unfortunately, the depairing electron correlations have not been estimated in the same way. For that reason, the values of the Coulomb pseudopotential have been chosen on the basis of the newest experimental data related to the critical temperature \cite{Karuzawa2002A} 
(see Table \tab{t1}). The Eliashberg equations on the imaginary axis have been solved for 1100 Matsubara frequencies ($M=1100$). We have taken the advantage of the numerical methods described and used in the works: \cite{Szczesniak2012N}, \cite{Szczesniak2013C}, \cite{Szczesniak2013E}, \cite{Drzazga2014A}, \cite{Szczesniak2014H}, and \cite{Szczesniak2015A}. The functions $\Delta_{n}$ and $Z_{n}$ are stable for the temperature higher than $T_{0}=1.5$ K. It is assumed that the cut-off energy is equal to $5\Omega_{\rm max}$, where the exact values of the Debye energy are collected in Table \tab{t1}.

\begin{table}[t]
\renewcommand*{\tablename}{\small{Tab.}}
\renewcommand*{\baselinestretch}{1}
\caption{\label{t1} The selected parameters determining the properties of the high-pressure superconducting state in phosphor.}
\begin{center}
\begin{tabular}{|c|c|c|c|c|c|}\hline
                        &                 &                   &                   &                    &                     \\
{\bf Quantity}  & {\bf Unit} &  {\bf $20$ GPa}   &  {\bf $30$ GPa}   &   {\bf $40$ GPa}   &   {\bf $70$ GPa}    \\
                        &                 &                   &                   &                    &                     \\
\hline
                        &           &                   &                   &                    &                     \\
$T_{C}$                 & K         & 6.39              &  8.45             &   8.05             &  5.4                \\ 
                        &           &                   &                   &                    &                     \\
$\Omega_{\rm max}$      & meV       & 59.4              &  62.3             &   64.8             &  74                 \\
                        &           &                   &                   &                    &                     \\
$\mu^{\star}$           &           & 0.37              &  0.29             &   0.27             &  0.28               \\
                        &           &                   &                   &                    &                     \\ 
$\lambda$               &           & 0.795             &  0.771            &   0.739            &  0.676              \\
                        &           &                   &                   &                    &                     \\
$\omega_{\rm ln}$       & meV       & 418.3             &  444.1            &   456.5            &  469.4              \\
                        &           &                   &                   &                    &                     \\
\hline
\end{tabular}
\end{center}
\end{table}

\begin{figure} 
\includegraphics[width=\columnwidth]{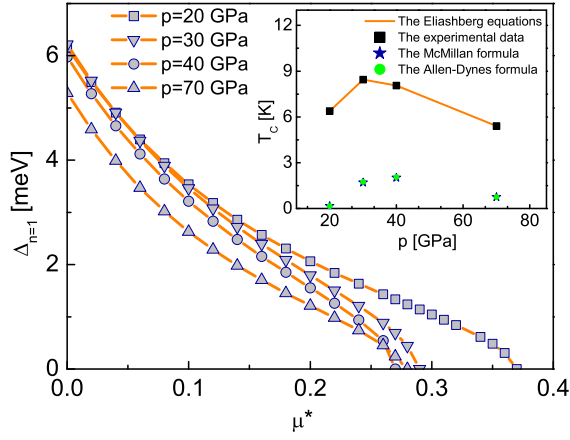}
\caption{\label{f2}
The maximum values of the order parameter as a function of the Coulomb pseudopotential. The exact values of the critical temperature have been collected in Table \tab{t1}. High values of $\mu^{\star}$ mean that the critical temperature cannot be estimated on the basis of the McMillan or Allen - Dynes expression \cite{McMillan1968A}, \cite{Allen1975A} (see the figure’s insert).}
\end{figure}

In the first step we have calculated the physical values of the Coulomb pseudopotential corresponding to the given pressure. For this purpose, the following condition has been used: $\left[\Delta_{n=1}\left(\mu^{\star}\right)\right]_{T=T_{C}}=0$, where $\Delta_{n=1}$ represents the maximum value of the order parameter. Figure \fig{f2} presents the dependence of the order parameter on the Coulomb pseudopotential at the critical temperature. The obtained results prove that $\mu^{\star}$ takes very high values (see Table \tab{t1}) in relation to the value generally taken into account in the calculations ($\mu^{\star}\sim 0.1$). It should be noted that this situation is often observed in the analysis of the high-pressure superconducting state. For example, for lithium the physical value of the Coulomb pseudopotential is equal to: $\left[\mu^{\star}\right]_{p=29.7 {\rm GPa}}=0.36$ \cite{Szczesniak2010A}. The anomalously high values of the Coulomb pseudopotential can be explained when analyzing the influence of the retardation effects on the value of the non-renormalized Coulomb pseudopotential ($\mu\equiv U\rho\left(0\right)$) in the second order of $\mu$, where 
$U\equiv\int\int d^{3}{\bf r}_{1}d^{3}{\bf r}_{2}|\Phi_{i}\left({\bf r}_{1}\right)|^{2}V_{C}\left({\bf r}_{1}-{\bf r}_{2}\right)|\Phi_{i}\left({\bf r}_{2}\right)|^{2}$, $\Phi_{i}\left({\bf r}\right)$ is the Wannier function. In the considered case the retardation effects lead to the reduction: $\mu\rightarrow \mu^{\star}$, but not as large \cite{Bauer2012A} as it was predicted by the classical Morel and Anderson theory, which was limited to the linear order with respect to $\mu$ \cite{Morel1962A}. 

\begin{figure} 
\includegraphics[width=\columnwidth]{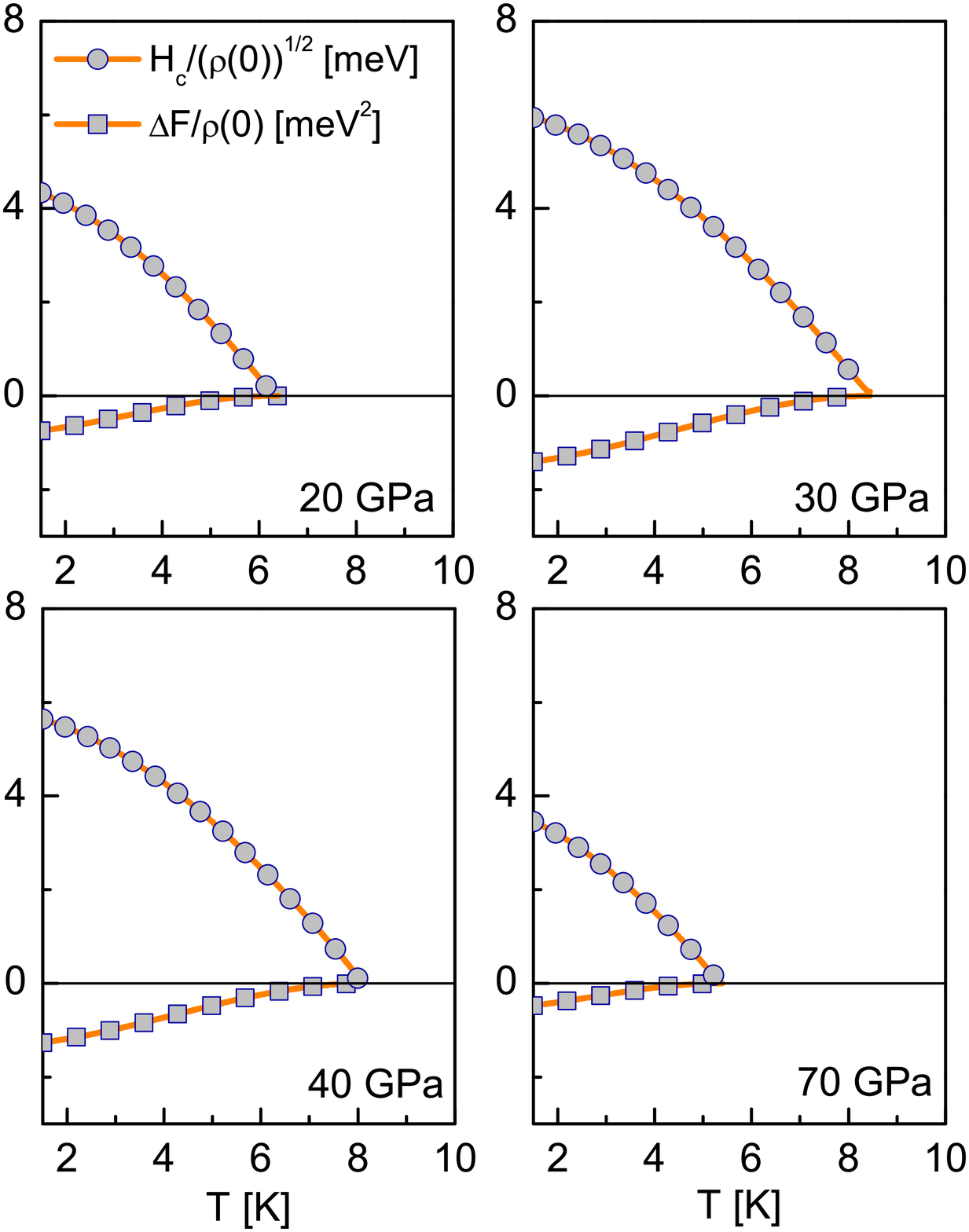}
\caption{\label{f3}
The free energy difference between the superconducting and the normal state as a function of the temperature (the lower panels). 
The thermodynamic critical field - the upper panels.}
\end{figure}
 
\begin{figure} 
\includegraphics[width=\columnwidth]{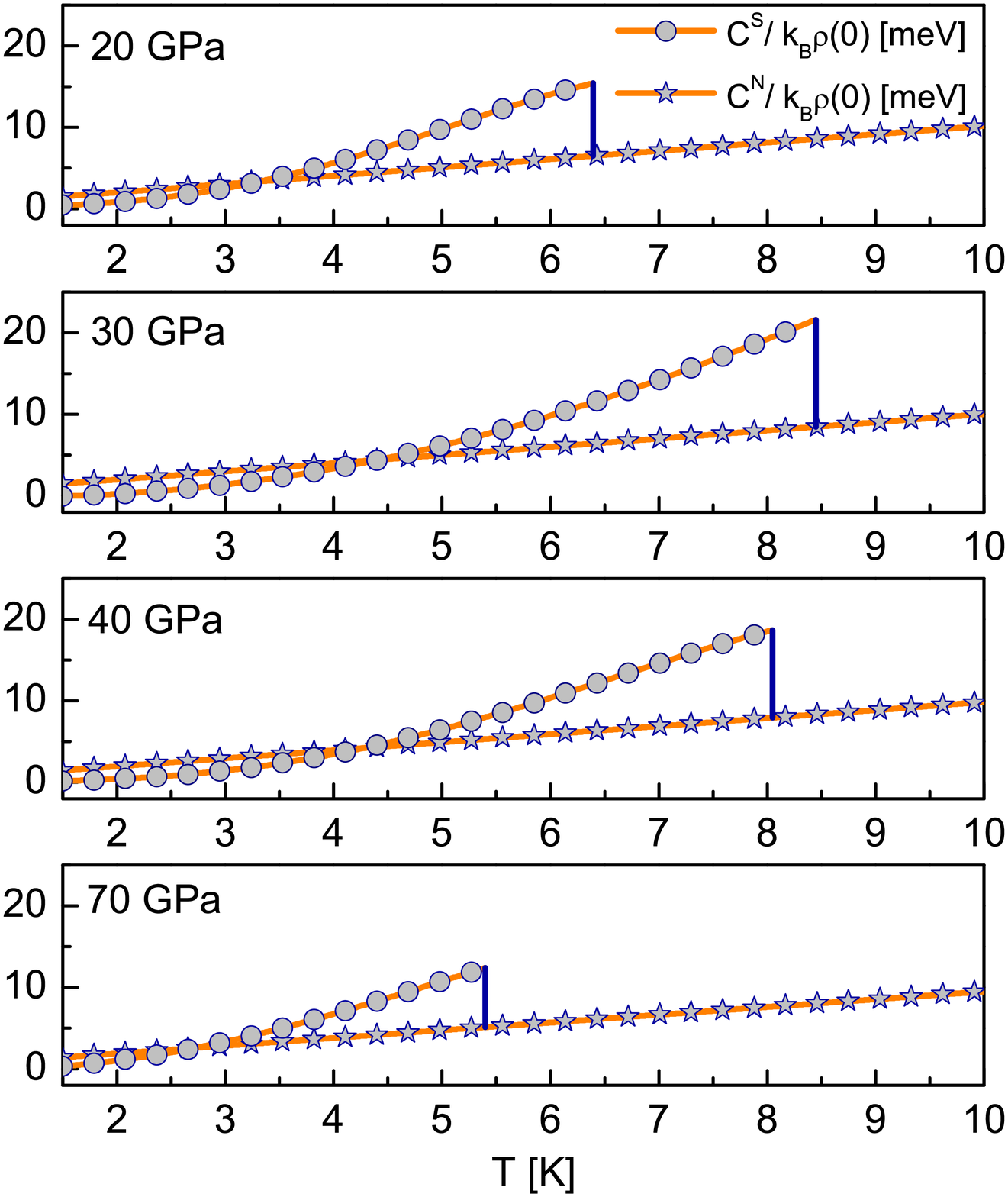}
\caption{\label{f4}
The specific heat of the superconducting state and the normal state as a function of the temperature.}
\end{figure}

In the Eliashberg formalism the free energy difference between the superconducting and the normal state should be calculated on the basis of the formula \cite{Bardeen1964A}:
\begin{eqnarray}
\label{r14}
\Delta F &=&-2\pi k_{B}T\rho\left(0\right)\sum^{M}_{n=1}
[\sqrt{\omega^2_n+\left(\Delta_n\right)^2}-|\omega_n|]\\ \nonumber
&\times&[Z^{\left(S\right)}_n-Z^{\left(N\right)}_n \frac{|\omega_n|}{\sqrt{\omega^2_n+\left(\Delta_n\right)^2}}].
\end{eqnarray}
In the next step, the thermodynamic critical field ($H_{C}=\sqrt{-8\pi\Delta F}$) has been
calculated, as well as the difference in the specific heat of the superconducting and the normal state: $\Delta C=C^{S}-C^{N}=-T\frac{d^2\Delta F}{dT^2}$, where: $C^{N}=\gamma{T}$. The Sommerfeld parameter is equal to: $\gamma\equiv \frac{2}{3}\pi^{2}k_{B}^{2}\rho(0)\left(1+\lambda\right)$. The electron - phonon coupling constant should be calculated on the basis of the formula: $\lambda\equiv 2\int^{\Omega_{\rm max}}_0 d\Omega \frac{\alpha^2F\left(\Omega\right)}{\Omega}$ (Table \tab{t1}). The lower panels in Figure \fig{f3} show the full form of the function $\Delta F\left(T\right)$. It can be very clearly seen that the free energy difference takes negative values in the whole range of the temperature, up to $T_{C}$, which is the evidence of the thermodynamic stability of the superconducting phase. It should be noted that in the case of phosphor the lowest value of the free energy difference has been obtained for $30$ GPa 
{($\Delta F\left(T_{0}\right)/\rho\left(0\right)=-1.4$ ${\rm meV^{2}}$)}, while the highest has been obtained for $70$ GPa 
{($\Delta F\left(T_{0}\right)/\rho\left(0\right)=-0.47$ ${\rm meV^{2}}$)}. From the physical side the obtained result is related to the values of the electron - phonon coupling constant and the Coulomb pseudopotential. The course of the function $\Delta F\left(T\right)$ directly determines the thermodynamic critical field and the specific heat. The results are presented in the upper panels of Figure \fig{f3} and Figure \fig{f4}, respectively.

\begin{figure} 
\includegraphics[width=\columnwidth]{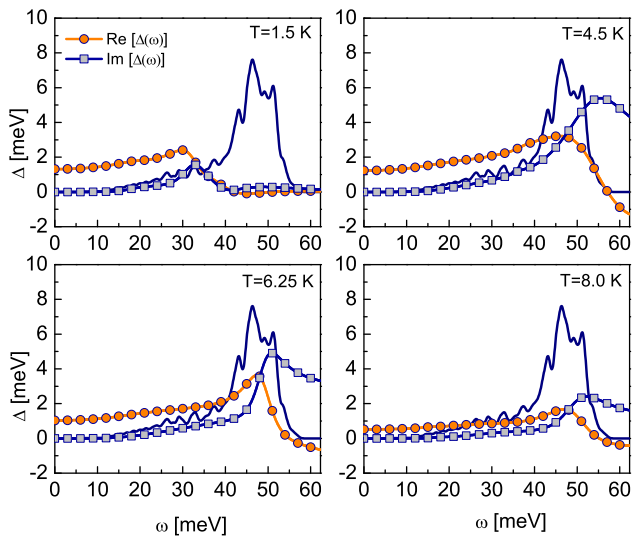}
\caption{\label{f5}
The order parameter on the real axis for the selected values of the temperature ($p=30$ GPa). Additionally, the shape of the rescaled Eliashberg function ($6\alpha^{2}F\left(\omega\right)$) was plotted.}
\end{figure}

The physical value of the order parameter should be calculated by using the equation:
$\Delta\left(T\right)={\rm Re}\left[\Delta\left(\omega=\Delta\left(T\right),T\right)\right]$, while the form of the order parameter on the real axis has been determined by solving the Eliashberg equations in the mixed representation. The exemplary courses have been collected in Figure \fig{f5}. When analyzing the behavior of the order parameter on the real axis, the attention should be paid to the fact that for low frequencies the non-zero values are taken only by the real part of the function $\Delta\left(\omega\right)$. From the physical point of view, this proves the existence of the infinitely long-lived Cooper pairs. At the higher frequencies the non-zero is also the imaginary part of the order parameter, which, of course, determines the finite lifetime of the electron pairs. At this point it is worth mentioning that the particularly large changes in the values of the order parameter are very closely correlated with the distinctive group of peaks occurring in the course of the Eliashberg function.   

%

\begin{figure} 
\includegraphics[width=\columnwidth]{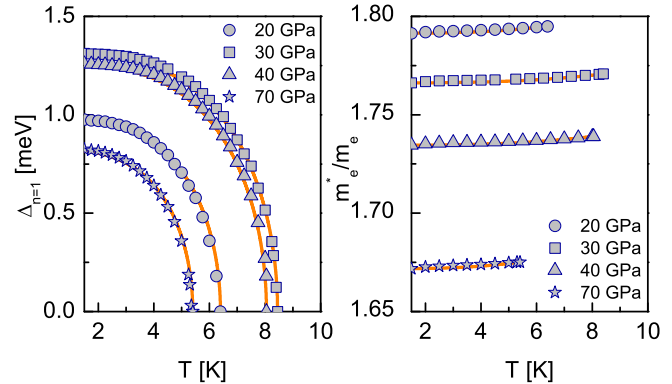}
\caption{\label{f6}
The dependence of the order parameter and the electron effective mass on the temperature.}
\end{figure}

Then, the electron effective mass ($m^{\star}_{e}$) has been calculated on the basis of the formula: 
$m^{\star}_{e}={\rm Re}\left[\left[Z\left(T\right)\right]_{\omega=0}\right]m_{e}$, where the symbol $m_{e}$ denotes the electron band mass. The values of the order parameter and the electron effective mass have been plotted in Figure \fig{f6}. 

The dimensionless ratios $R_{\Delta}$, $R_{C}$, and $R_{H}$ have been calculated in the last step. The obtained results prove that the values do not significantly differ from the values predicted by the BCS theory. The biggest derogations of several percent have been found for the pressure at $30$ GPa. Let us note that the result above is related to the insignificant strong - coupling and retardation effects, which are characterized by the ratio: 
$r\equiv k_{B}T_{C}/\omega_{\ln}$. Within the of BCS, the Eliashberg equations predict: $r\rightarrow 0$. For $p=30$ GPa, it was obtained: $r=0.02$. The quantity $\omega_{\ln}$ is called the logarithmic frequency, and should be determined with the help of the formula: 
$\omega_{\ln}\equiv \exp\left[\frac{2}{\lambda}\int^{\Omega_{\rm max}}_{0}d\Omega\frac{\alpha^{2}F\left(\Omega\right)}{\Omega}\ln\left(\Omega\right)\right]$. The values of the logarithmic frequency have been collected in Table \tab{t1}.

\section{Summary}

The presented work discusses the Eliashberg formalism, which is used for the quantitative description of the thermodynamic properties of the superconducting condensate induced by the electron - phonon interaction. The detailed considerations have been illustrated on the example of the superconducting state in phosphor under the influence of the high pressure. With respect to the considered system it has been found that the critical temperature is relatively low, which is connected with not very high values of the electron - phonon coupling constant and the significant depairing Coulomb interactions. Additionally, it has been shown that the strong - coupling and retardation effects do not cause the significant derogations of the values of the dimensionless ratios $R_{\Delta}$, $R_{C}$, and $ R_{H}$ from the values predicted by the BCS theory.

Let us point our attention toward the fact that the Eliashberg formalism discussed in the presented work is the main tool used for the description of the high - temperature superconducting state induced by the electron - phonon. In particular, it can be used for the quantitative analysis of the superconducting state in ${\rm H_{2}S}$ and ${\rm H_{3}S}$ \cite{Drozdov2015A}, \cite{Durajski2015C}, \cite{Duan2015A}, \cite{Durajski2015D}. 
 
\bibliography{template}
\end{document}